\documentclass{article}

\usepackage{arxiv}

\usepackage[utf8]{inputenc} % allow utf-8 input
\usepackage[T1]{fontenc}    % use 8-bit T1 fonts
\usepackage{natbib}
\usepackage{hyperref}       % hyperlinks
\usepackage{url}            % simple URL typesetting
\usepackage{booktabs}       % professional-quality tables
\usepackage{amsfonts}       % blackboard math symbols
\usepackage{nicefrac}       % compact symbols for 1/2, etc.
\usepackage{microtype}      % microtypography
\usepackage{lipsum}
\usepackage{graphicx}

\title{Towards Semi-Automatically Comparing Keyword-Based and Semantic Search Accuracy}

\author{
  Mohamed Ben Salha \\
  Technical University of Munich \\
  Heilbronn, Germany \\
  \texttt{mohamed.bensalha@tum.de} \\
  \And
  Fiete Lüer \\
  Gofore GmbH \\
  Munich, Germany \\
  \texttt{fiete.lueer@gofore.com} \\
  \AND
  Maik Betka \\
  University of Stuttgart \\
  Stuttgart, Germany \\
  \texttt{maik.betka@iste.uni-stuttgart.de} \\
  \And
  Stefan Wagner \\
  Technical University of Munich \\
  Heilbronn, Germany \\
  \texttt{stefan.wagner@tum.de} \\
}

\begin{document}
\maketitle
\begin{abstract}
The increasing importance of Information Retrieval (IR) in managing large datasets has highlighted significant limitations in traditional keyword-based search systems. 
Context-aware chat-based search methods, such as Retrieval Augmented Generation (RAG), have recently emerged, but their evaluation compared to keyword-based systems often relies on subjective user feedback. 
A rigorous, quantitative comparison between these paradigms remains lacking. 
This work introduces a novel, preliminary framework to quantitatively assess IR accuracy of search systems that produce different output formats, such as lists and messages.
It focuses on two key aspects: the ranking accuracy for keyword-based systems and the completeness of retrieved information for semantic chat-based systems.
Our approach enables semi-automatic comparisons of semantic and keyword-based methods using interchangeable equivalence classes tailored to domain-specific contexts (e.g., companies or problems).
We validate the framework through an industrial case study, demonstrating statistically significant improvements in context-aware search over keyword-based methods, supported by analyses including the Mann-Whitney U-Test. 
With its adaptable design, the proposed framework provides a strong foundation for objectively assessing keyword-based and semantic chat-based search methods.
% Maik: Ich würde noch betonen, dass der Vergleich darauf beruht, dass du ja das eigentliche Ziel ("wie gut finde ich was ich suche") als Grundlage dafür nimmst zu bewerten wie gut beide Systeme im Vergleich sind. Bei search engines / listenplatzierten Systemen weiß man, dass die Platzierung der Suchergebnisse eine Rolle spielt, bei RAGs sollten die Informationen gut im Text zusammengefasst sein. Das ist die Grundlage für den Vergleich. Das geht leider etwas im Abstract unter.
%Fiete: Was ich geändert habe: ich habe den Fokus etwas von user-friendly und chat-based genommen. Beides ist für uns inhaltlich irrelevant und am Ende vergleichen wir v.a. auch Methoden bei denen user-friendliness oder chat egal ist. Habe dafür den Fokus eher auf den context-aware bzw semantic Teil gelegt. Wollen wir am Ende noch nen Satz dazu sagen wie Metriken wie CTR dort reinspielen und was das Framework dahingehend ermöglicht?. Was an sich immer schön ist, ist wenn man im abstract auch noch 1-2 für typische Leser bekannte Kennzahlen nennen kann, die die Signifikanz der Messungen (bspw bei RAG vs keyword die Signifikanz basierend auf den Tests) unterstreichen. Das aber nur wenn es im entsprechenden Bereich bekanntere Metriken sind. Insgesamt wäre es aber gut mehr als "better performance" (habe ich jetzt auch nochmal umschrieben) zu sagen, irgendwas das die praktische Relevanz unterstreicht.
\end{abstract}

%% the work being presented. Separate the keywords with commas.<<
%\keywords{Do, Not, Use, This, Code, Put, the, Correct, Terms, for,
%  Your, Paper}


\maketitle

\section{Introduction}
% Key arguments:
% - IR is necessary and keyword based approaches has disadvantages
% - LLMs are up and coming but hallucinate and do not have access to internal knowledge
% - RAGs are better and can be used to retrieve semantic information but there is no systematic /objective quantification of the goodness in comparison to keyword based approaches before actually implementing it and measuring e.g. the actual CTR
% How can we quantify and compare different approaches to estimate if information is found (before actually implementing it to avoid costs/disruption?
% - We propose a framework used to evaluate different IR methods using exchangable metrics, allowing finetuning the actual performance to problem and company specific results (was ich meine: Neue Confluenceversionen stellen die Ergebnisse anders dar als alte Versionen, das Framework erlaubt die Anpassung der Absprungwerte basierend auf Präferenz (wie hart bin ich bei erlaubter CTR) und vorher gemessener, generalisierter CTR. 
In today's data-driven world, organizations and individuals rely heavily on vast repositories of documents and wikis to store and access information.
However, the growth in size and complexity of these collections makes efficient Information Retrieval (IR) challenging. 
Traditional keyword-based search methods often fall short in understanding context and handling typographical errors~\cite{search_engine_beyond_keyword_search}. 
In recent years, advancements in Natural Language Processing (NLP) have paved the way for more sophisticated IR methods. 
One of the most important developments is the emergence of RAG~\cite{rag_for_nlp}, which harnesses Large Language Models' (LLMs) text understanding and generation capabilities by incorporating an external retrieval mechanism.
It provides the answer-generating LLM with retrieved context, mostly not part of the training corpus of that LLM, enabling more context-aware and accurate responses to user queries and reducing the likelihood of hallucinations.

%Unlike LLMs, RAG mitigates issues such as hallucinations by grounding its responses in relevant documents semantically retrieved from a preprocessed corpus. 
%Fiete: Aber RAGs können im Generation teil ja auch hallucinaten. 
% Eher: The retrieval mechanism allows to utilize information unknown to the LLM during training and offering it as context to the LLM generation. This reduces the likelihood of hallucinations and allows for more context-aware and accurate responses to user queries (oder so) -> done
%As a result, RAG allows for more context-aware and accurate responses to user queries.
While RAG is meant to offer better retrieval performance compared to traditional keyword-based search, a direct quantitative comparison of both methods is still required and missing for a more complete evaluation.
%This work addresses this gap by conducting an industrial case study.
%Fiete: Die contribution is keine case study, die study ist nur ein Mittel zur Verifikation. Im Folgenden ein Vorschlag um das umzuschreiben (inhaltlich nicht wirklich anders nur teilweise anders umschrieben)
%Eher: -> übernommen
This work addresses this gap by investigating how the correctness and completeness of information can be used to quantitatively measure the accuracy of a context-aware search compared to a keyword-based one. 
We propose a preliminary framework to semi-automatically evaluate different IR methods using interchangeable equivalence classes, allowing the fine-tuning of the actual performance to problem and company-specific results. 
This is validated in an industrial case study, where the IR accuracy of the search engine of Confluence\footnote{\url{https://www.atlassian.com/software/confluence}}, a widely used keyword-based wiki, is compared with RAG.
We provide supplementary materials, including source code and resources, to facilitate the reproduction of our results\footnote{\url{https://doi.org/10.5281/zenodo.23040515}}.
% Maik: Quelle fehlt/funktioniert nicht (github_anonymous).   -> momentan nur platzhalter, wird noch ergänzt.
\section{Goal}

On the one hand, with an input of keywords, the output of keyword-based search engines has the format of a vertical list of results, allowing for the display of several potentially relevant entries. 
The higher the position of an entry is, the higher its relevance should be~\cite{search_engine_beyond_keyword_search}. 
Click-Through Rate (CTR) is a metric that expresses the percentage of users who click on a particular result after viewing it in the search results list~\cite{ctr}.
This metric demonstrates that users' clicks on search results drop off exponentially the lower a result appears on the page.

On the other hand, the emergence of chat-based search systems and those capable of semantic search (e.g., RAG) has marked a significant shift in the field of IR.
Taking an input mostly in natural language, they generate text-based outputs that are evaluated on completeness, hallucination, correctness, and faithfulness~\cite{rag_evaluation_survey,rag_evaluation_metrics_and_datasets}.

\begin{figure}[!ht]
    \centering
    \includegraphics[width=1\linewidth]{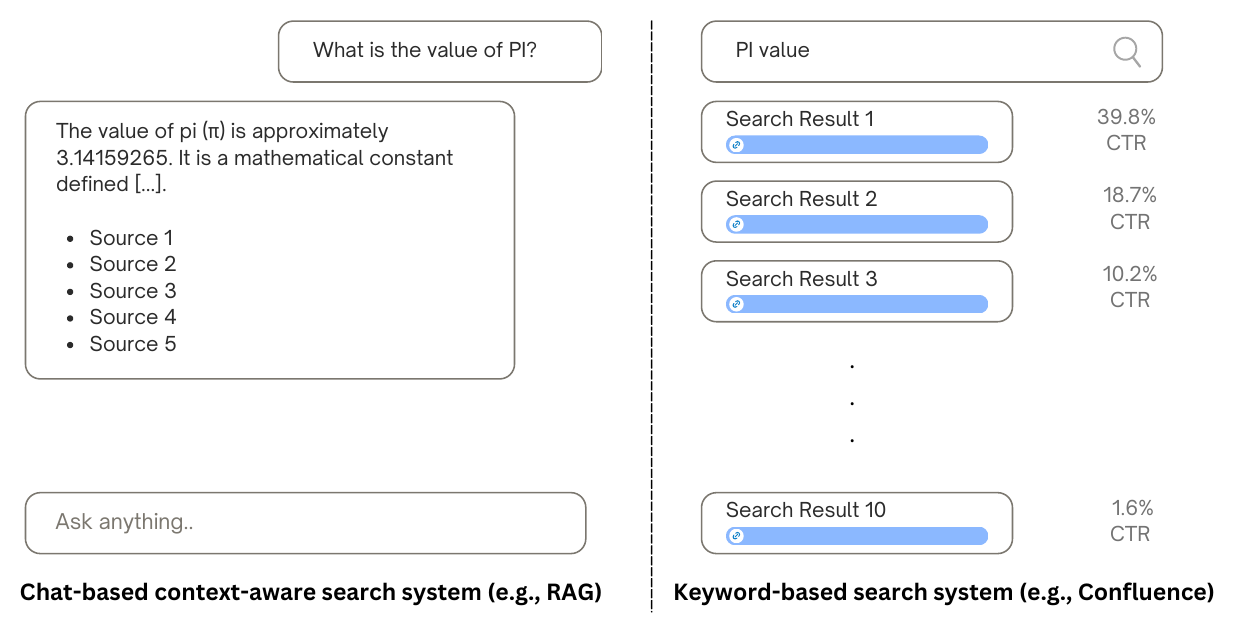}
    \caption{Chat-based and keyword-based search approaches with their different output formats.}
    \label{fig:search_aproaches}
\end{figure}

Figure~\ref{fig:search_aproaches} demonstrates the different output formats of both these search systems and the corresponding CTR for each entry for list-based search results~\cite{ctr}.

Given that both IR systems access the same data, the question of which one delivers more adequate results for users' queries is both compelling and requires a thorough investigation.
However, the different output formats make a direct comparison of their quality without making further adjustments difficult.
Our aim is to tackle this challenge by proposing an evaluation framework.

\section{Related Work}
\cite{related_work_comparison_keyword_and_semantic_search_engines_2014} presented a comparative performance evaluation between three semantic web-based and two keyword-based search engines.
The authors use an ontology-based crawler to resolve the problem of synonyms and polysemy for semantic search.
The evaluation dataset contained 10 queries from various topics.
The first twenty documents on each retrieval output were classified using human judgment as being “relevant” or “non-relevant” to calculate the precision in a further step.
The authors find that semantic web-based search engines perform better than keyword-based ones.
% Maik: Was sind diese Werte? Wenn sie nicht zwingend notwendig sind würde ich den Nebensatz weglassen. 
%-> das sind P@20, precision von den ersten 20 ergebnissen, wir hatten ja damals P@5. wurden erwähnt, um zu zeigen, wie besser die waren.
To conclude their work, \cite{related_work_comparison_keyword_and_semantic_search_engines_2014} presented the advantages and limitations of both search approaches in terms of input and output relevancy.
This study compares semantic and keyword-based search engines that provide search results as a list. 
However, it does not compare these systems to other systems that produce different results, such as the conversational messages generated by chat-based systems (e.g., RAG) compared to traditional keyword lists. 
\cite{related_work_comparison_keyword_and_semantic_search_engines_2014_with_other_queries_and_search_engines} conducted a similar experiment, but considered other search engines and queries from the medical domain.
%Interestingly, while both studies used the same methodology and were performed in 2014, they had a dichotomous classification of Google and Yahoo as semantic web search engines by the first study and as keyword-based ones by the second study, and DuckDuckGo the other way around.
%Fiete: den vorherigen Satz kann man bei Platzproblemen auch weglassen - ist für die Evaluation eher irrelevant
%This highlights the lack of clarity in search engine features and the need for explicit specifications.
%\citeauthor{related_word_conversional_vs_traditional} \cite{related_word_conversional_vs_traditional} performed a comparative study of user behavior and outcomes in traditional versus conversational legal case retrieval systems. The study employed legal experts as an intermediary agent to simulate the conversational search process. The authors gathered feedback in terms of user satisfaction, interaction, and the perceived workload. Conversational search achieved a higher search success rate and led to an overall better user experience. However, the conversational search, simulated by humans, overshadows the actual capabilities of AI and raises concerns about its applicability in such domains. It further introduces a source of bias because users assume they interact with an AI agent. 
\cite{related_work_tu_darmstadt_scientific} developed a conversational agent to help NLP researchers efficiently explore literature.
For natural language understanding, they use both a rule-based and a machine learning-based approach.
The authors performed a user-based evaluation to assess usage aspects such as diversity and information coverage, achieving satisfactory results.
However, user-based evaluations have several challenges.
They are time-consuming, costly, and the subjective results may be biased due to an unrepresentative sample~\cite{qualitative_quantitative_methods_advantages_and_disadvantages}.
%Fiete: Hier noch auf die Nachteile von user based evals eingehen (subjektivität, Kosten?)

%To overcome the challenges traditional keyword-based systems have, extensive work in the domains of semantic and conversational search has been done.
%Nevertheless, an objective evaluation and comparison of these systems producing different output formats remains lacking~\cite{related_work_macaw,related_work_scholar}. 
Despite extensive work in the domains of semantic and conversational search to overcome the challenges traditional keyword-based systems have, an objective evaluation and comparison of these systems producing different output formats remains lacking~\cite{related_work_macaw,related_work_scholar}. 
\section{Evaluation Framework}

The core of our evaluation framework is the quantitative comparison of different IR approaches despite their different output formats~(e.g., list, message).
Both the correctness of the results (retrieved results should be relevant) and the ranking (the most relevant result should be on top of the list) need to be accounted for. 
To achieve this, we perform the following steps:
\begin{enumerate}
    \item Creation of a ground truth to account for language peculiarities (e.g., semantic equivalence and contextual reasoning). The ground truth is created by defining a checklist of what content should be covered in the search result and creating queries where some parts are altered, leading to queries with: (1) exact keywords (to evaluate direct matching of the terms used in the documents and the search), (2) synonyms (to evaluate semantic equivalence), and (3) synonym-based queries on unrelated topics (to evaluate contextual reasoning). Having an evaluation dataset spanning these categories better reflects the usage of IR systems because it covers different aspects such as complexities, writing styles, and lengths.
    \item Creation of task-specific equivalence classes containing outputs that are considered qualitatively similar. These can be used to derive an ordinal scale whose levels can differ according to the problem, allowing a comparison of approaches. As such, it is necessary that the scale allows for a clear ranking (e.g., "All relevant information is retrieved", "Some relevant information is retrieved", and "No relevant information is retrieved"). Users are free to define the equivalence classes and which ones to assign to which results.
     \item Evaluation of the quality of the IR approaches via statistical tests (e.g., Mann-Whitney U-Test and Vargha-Delaney A12), to assess statistically significant differences between the two approaches and measure their effect size.
\end{enumerate}
%This emphasizes the flexibility and adaptiveness of the evaluation framework and enables users to define custom categories and metrics based on the targeted retrieved content quality and its differentiations. 
%Ende Vorschlag
\section{Experimental Setup and Results}
% Key aspects:
%- industrial partner using confluence/lucine
% - Skala 1-3 
% - CTR als Approximation da relativ generalisiert, übertragbar und erwartet stabil über Zeit/Versionen (nutzer werden eher weniger anfangen Folgeseiten anzuklicken
% RAGs leicht in das Framework integrierbar. Warum RAG? Ein Schritt weniger für Nutzer (muss Ergebnis nicht anklicken sondern bekommt Info direkt). Key finding: RAGs schlechter als retrieval -> was ist der tradeoff/Vorteil?
% Vergleich groundtruth mit erstelltem CTR score und nutzerstudie?
To test the applicability of our preliminary framework in a real-world setting, our research adopts an industrial case study approach where Confluence is used as a keyword-based wiki.
We implemented a RAG application with \textit{ChromaDB} as a vector database, \textit{Llama 3.2 8B} as a LLM for the generation part, and \textit{jinaai/jina-embeddings-v2-base-de} as an embedding model, as the Confluence data is in German. 
For chunking, we do not rely on the conventional method that uses a predefined length. 
Instead, we chunk the document content based on its hierarchical structure to ensure the retention of related information within each chunk.
The retrieval component uses a \textit{K}-based approach, where \textit{K} specifies the number of elements to be retrieved based on their cosine similarity to the query~\cite{retrieval_quality_rag_evaluation}. 
In this work, we set $K=5$.

While this paper focuses merely on the evaluation framework, we also conducted semi-structured interviews as part of the broader research industrial case study to assess users' perception of the Confluence search engine.
Therefore, the interviews are not described or discussed in detail here. 
However, we asked interviewees to create a diverse set of example queries, in keywords and natural language, to reflect real-usage data.
We then slightly adjusted these queries and paraphrased each one three times to cover variations across three critical dimensions: complexity, language, and length, resulting in a total of 63 queries for each IR approach.
With that being done, we ensure the validation of our evaluation framework with two different systems. 
We first compare Confluence with the entire RAG pipeline, as this provides an explicitly summarized answer matching the query (so the user does not have to read the entire article). 
However, this comes with a potential risk of missing or incorrect information (hallucination).
Next, we compare Confluence with only the retrieval component of RAG, as it represents the semantic search part, excluding hallucination possibilities.

With the creation of the ground truth, the first step of the framework is completed.
The second step encompasses the creation of equivalence classes to group outputs with a similar retrieval quality.

\subsection{Confluence vs. RAG} 
In this case, the industrial partner relies on three equivalence classes and uses an ordinal scale of three levels: \{1,2,3\} (lower value is better IR accuracy), signaling if all, some, or no expected information is present. 
The Confluence search returns multiple results in a list format, allowing for the display of several potentially relevant articles. 
However, the most relevant entries (i.e., all information is present) are expected to be more salient. 
%To account for the relevancy, we integrate the expected user behavior.
%Click-Through Rate (CTR) is a metric that expresses the percentage of user interactions to the number of times a result is displayed~\cite{ctr}. 
Referring to CTR allows us to reward saliency (low value/equivalence class) and punish missing or non-salient information (high value/equivalence class).
%Based on \cite{ctr}, the likelihood of users clicking on search results drops off exponentially the lower a result appears on the page.
% Being in the first two positions ensures optimal visibility and captures 60\% of CTR, where the complete information should be listed as well.
% From the third to seventh ranking position, the moderately visible rest captures 35\% CTR.
% In the context of RAG, partially correct answers are expected to fall under this second category.
% Results ranked eighth or higher, or those not displayed at all, collectively attract only about 5\% of clicks, indicating minimal interaction.
% The RAG counterpart to this is an output missing relevant information (e.g., only hallucination).
Table~\ref{tab:evaluation_framework_conf_vs_rag} shows an overview of this scoring schema.
This grouping reflects a discretionary decision, made in consultation with the case company, based on practical considerations relevant to the case.
\begin{table}[!h]
\centering
 % \resizebox{\linewidth}{!}{%
\begin{tabular}{ccc}
\toprule
\textbf{Value} & \textbf{Confluence (CTR)} & \textbf{RAG (via checklist)}  \\
\midrule
1 & 1 - 2 (60\%) & Information is fully complete   \\
2 & 3 - 7 (35\%) & Information is partially complete  \\
3 & 8 - 10 (5\%) & Information is unavailable \\
\bottomrule \\
\end{tabular}
%}
\caption{Evaluation framework for Confluence vs. RAG IR accuracy/information availability. Each row represents an equivalence class assumed in this work. Lower values indicate better IR accuracy.}
\label{tab:evaluation_framework_conf_vs_rag}
\end{table}

\begin{figure}[!h]
  \centering
   \includegraphics[width=\linewidth]{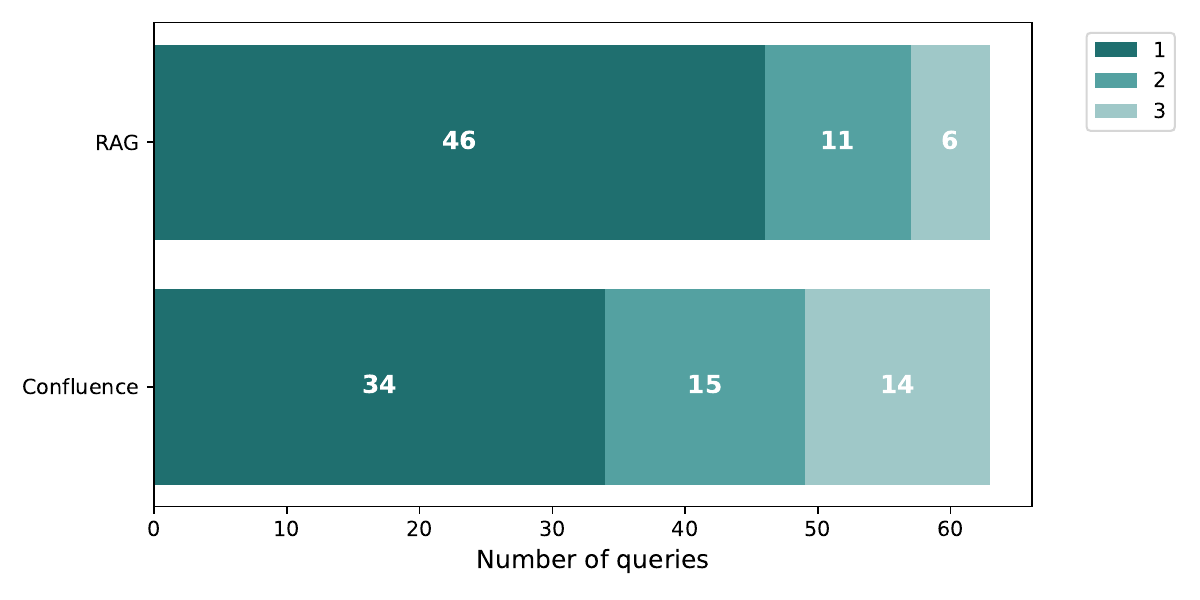}
  \caption{IR accuracy comparison between Confluence and RAG on a dataset of 63 queries. Lower values indicate better IR accuracy.}
  \label{fig:conf_vs_rag_bar_plot}
\end{figure}

Having established the equivalence classes set, we proceed with the third step of our evaluation framework, namely the evaluation of the IR quality by means of statistical tests.
To quantify the difference in performance between groups, we use the $A12$ statistic, a non-parametric effect size measure proposed by Vargha and Delaney~\cite{vargha-delaney-a12}. 
%In the research context of this paper, given a performance measure M, the $A12$ statistic measures the probability that randomly performing a search query by the first group yields higher M values (i.e., lower performance) than performing a search with the second group. %$A12 > 0.5$ indicates a tendency towards the first group, $A12 < 0.5$ favors the second group, and in case of equivalence,  $A12 = 0.5$ ~\cite{statistical-tests-random-algorithms-se}. 
It quantifies in how many cases system 1 achieved a higher value than system 2. 
Since a higher equivalence class value signifies a worse performance, a $A12$ statistic of 0.397 indicates that in approximately 40\% of the evaluated queries, RAG achieved higher values, denoting a worse performance.
Thus, in 60\% of evaluated cases, RAG had better performance compared to the Confluence search engine regarding IR.
Figure \ref{fig:conf_vs_rag_bar_plot} depicts the aggregated performance among all queries and underpins the advantages of RAG regarding IR.

\subsection{Confluence vs. RAG Retrieval Component}

We also compare only the retrieval component to Confluence, because it represents the semantic search part of RAG without the generating component that possibly causes hallucination.
Furthermore, it displays retrieved content in a ranked list, allowing for a comparison with Confluence based on results positions, where each one represents an equivalence class.
%In this vein, we also adjust the evaluation framework to the specific needs of both systems by using an ordinal scale from 1 to 10, with 1 indicating the best value and 10 the worst. 
% Table~\ref{tab:evaluation_framework_conf_vs_retrieval} shows an overview of this scoring schema.
% \begin{table}[!h]
%     \resizebox{\linewidth}{!}{%
% \begin{tabular}{ccc}
% \toprule
% \textbf{Value} & \textbf{Confluence} & \textbf{Retrieval} \\
% \midrule
% Actual position number & 1 - 10 &  1 - 5 \\
% 10 & > 10 or not found   & not found in top 5 \\
% \bottomrule
% \end{tabular}
% }
% \caption{Evaluation framework for Confluence vs. Retrieval Component of RAG IR accuracy. Lower values indicate better IR accuracy.}
% \label{tab:evaluation_framework_conf_vs_retrieval}
% \end{table}
% Maik: Die Tabelle ist etwas verwirrend, evtl. kannst du das im Fließtext machen und etwas Platz sparen. Dafür das "Aim/Objective" Topic nach der Introduction einbauen wie abgesprochen.
%For Confluence, search results within the top 10 positions are assigned a value equal to their position number.Results ranked beyond the 10th position receive a default value of 10. Similarly, for retrieval, results ranked within the top 5 positions are given a value corresponding to their position number, while missing ones are assigned a value of 10.
For Confluence, search results within the top 10 positions are assigned an equivalence class equal to their position number, while those beyond 10 receive a default value of 10. For retrieval, the top 5 results are assigned an equivalence class equal to their position, while missing ones are assigned a default value of 10.
This stricter scoring reflects the higher expectation for retrieval accuracy when using a smaller subset.
It emphasizes the importance of ranking the most relevant documents at the top to ensure effective IR. 

\begin{figure}[!h]
  \centering
  \includegraphics[width=\linewidth]{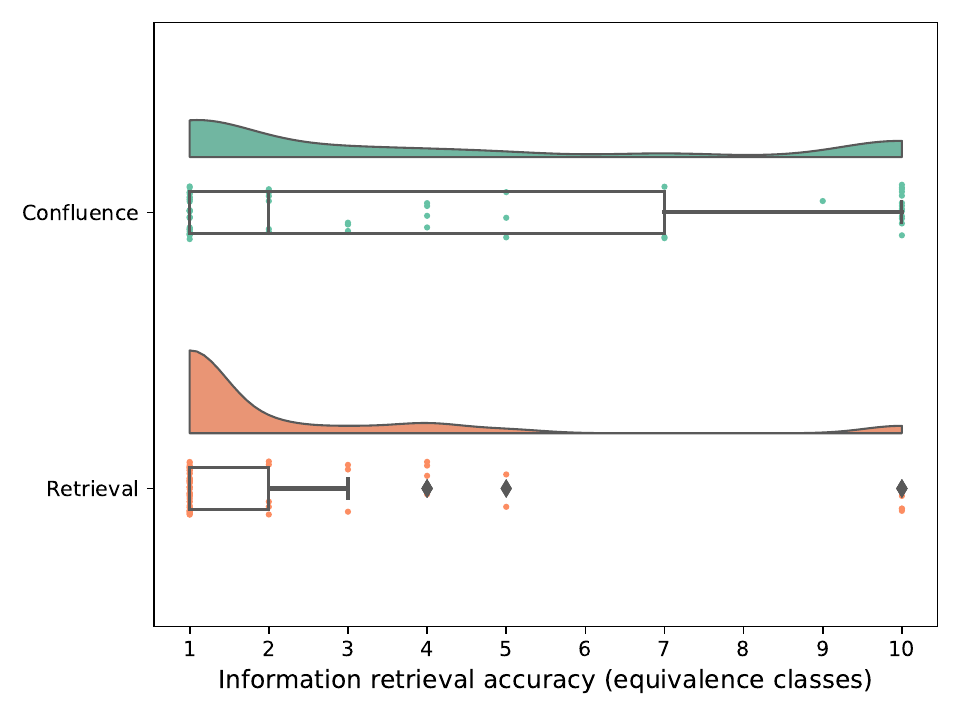}
  \caption{IR accuracy comparison between Confluence and the Retrieval part of RAG on a dataset of 63 queries. Lower values indicate better IR accuracy.}
  \label{fig:conf_vs_retrieval_raincloud_plot}
\end{figure}

The data presented in Figure \ref{fig:conf_vs_retrieval_raincloud_plot} substantiates the more accurate IR of the retrieval component of RAG compared to Confluence. 
% Both approaches show left-skewed distributions, but the retrieval component is more concentrated around the optimal value with fewer outliers.
% Its interquartile range (IQR) is narrower at 1 compared to Confluence's $IQR$ of 6, indicating more consistent and higher-performing results. 
% In terms of upper quartiles, the retrieval component has a $Q3$ of 2, while Confluence's $Q3$ is 7.
Considering that lower values indicate better IR accuracy, the data distribution and the box-plots corroborate the Vargha-Delaney $A12$ statistic of 0.353, shown in Table \ref{tab:mann_whitney_results}
This comparison is another validation of the preliminary framework and emphasizes its adaptiveness to different IR systems. 

\subsection{Statistical Tests}

Table \ref{tab:mann_whitney_results} presents the results of the Mann-Whitney U-Test and the Vargha-Delaney $A12$ statistic. 
The Mann-Whitney U-Test, or Wilcoxon rank-sum test, is a non-parametric statistical test used to identify statistically significant differences between the distributions of two independent samples. 
Unlike the T-Test, it does not assume normality and relies on the ranking of values~\cite{mann-whitney-u-test}.
%The Mann-Whitney U-Test, also known as the Wilcoxon rank-sum test, is a non-parametric statistical test used to determine whether there is a significant difference between the distributions of two independent samples. Unlike the T-Test, it does not assume normality and is based on the ranking of values~\cite{mann-whitney-u-test}.
%It assesses whether one of the two samples tends to have larger values than the other by ranking all observations from both groups together and then comparing the sum of ranks between the groups.
%Fiete: ich find die Erklärung an sich gut weil ich mit der Thematik sonst weniger zu tun habe, falls das in dem Bereich gängig ist, dann kann das hier verkürzt werden
We observe statistically significant differences between both approaches, respectively (p<0.05 threshold). 

%Fiete: Würde den satz vorher weglassen wenn wir wenig platz haben, hat nicht viel inhalt und ist klar denk ich -> done

\begin{table}[!h]
    \centering
%    \resizebox{\linewidth}{!}{%
        \begin{tabular}{cccccc}
            \hline
            \textbf{Group 1} & \textbf{Group 2} & \textbf{U Statistic} & \textbf{p-value} & \textbf{Effect Size} & \textbf{A12 Statistic}\\
            \hline
            RAG & Confluence & 1574.5 & 0.0195 & 0.1782  & 0.397  \\
            Retrieval & Confluence & 1400.5 & 0.0015 & 0.2538  & 0.353\\
            \hline \\
        \end{tabular}
  %  }
    \caption{Results of Mann-Whitney U-Test and Vargha-Delaney $A12$ statistic with a dataset of 63 entries for each group.}
    \label{tab:mann_whitney_results}
\end{table}

\section{Discussion}
We showcase the adaptiveness of the framework and its applicability to different IR systems. 
However, we note threats to validity, which open up avenues for future work.

% Notes:
% - We can also use expand this and easily check if different IR algorithms/used tools lead to changes in the CTR
% Key points:
% - CTR übertragbarkeit angenommen von Google zu Confluence. In der Praxis wird das anders sein aber die Schwankung der CTR wird nicht so riesig sein da nicht mehr als 10 Ergebnisse auf der ersten seite angezeigt werden. CTR heißt nicht dass es Angeklickt wird aber, dass es sichtbar ist und tendenziellm angeklickt werden kann
% - Vergleich möglich mit LLM as a judge (sind alle Infos da?) -> automatisierterer Vergleich ohne groundtruth mittelfristig, aber LLM abhängig
% Outlook: Mehr Engineering rund um RAGS, um eine Antwort zu verifizieren (E.g fetch actual text with semantic relevance from database)

%- evaluation still requires manual work, but it can be reduced 
%- Retrieval better. , since without generation part less energy consumption and provides direct sources

\subsection{Threats to Validity}

For our definition of IR accuracy, the focus is only on finding the desired information with a disregard for other aspects (response time, interactions, etc).
Second, we assumed the transferability of CTR from Google~\cite{ctr} to Confluence.
Although variations are expected, we assume these are negligible since, similarly, no more than ten results are displayed on the results page, and higher-ranked search results are expected to be more clicked~\cite{search_engine_beyond_keyword_search}.
% Moreover, CTR is commonly used to express how users engage with page content.
% However, our analysis primarily focused on the position of the most relevant result and assumed users would click it.
% This approach does not directly reflect users' actual behavior, but is used as an estimation.
% It is also likely that the CTR of the first results is also high because the results are usually of high quality in established search engines (correlation/causation). 
% However, the assumptions imposed by the selected metric can be easily exchanged in our framework, making it also flexible to handle different environments. 
Adaptions across domains are limited by the usage of different tools (with different CTRs) as well as the desired goodness and resulting ranking. 
Ranking the queries can be difficult in systems where, e.g., many results fulfill all aspects of a correct result, and setting good cut-off values for equivalence classes can be difficult in dynamic systems with growing documentation. 
On the other hand, the CTR allows an up-front estimation of the goodness of different IR algorithms even before their actual implementation in production environments.

The manual creation of the evaluation dataset may inadvertently leave out important entries or edge cases that are crucial for the IR systems' performance in production.
However, we ensured to cover a broad spectrum of the underlying data by involving employees from the case company.

The validation of our framework was performed through an industrial case study at only one company.
Hence, the results are not necessarily generalizable to other specific industrial environments and settings.
However, this work demonstrated a proof of concept and can be extended to further companies.

\subsection{Future Work}
Future work can tackle these challenges. By creating a freely available ground truth with well-defined contextual areas, a stricter control over the assumptions is given. This would also open up reproducible and comparable evaluations (without the need for repeated user studies).
Setting up the ground truth can be automated more extensively using current means (e.g., generating semantically similar queries or automatically creating preliminary checklists for completeness/correctness or prelabeling of information based on the checklist). Since means to automate this process are biased, a manual control of the generated ground truth is necessary.

%It should enable automatic score assignment based on the search output of both approaches to obviate the need for manual and repetitive tasks.
Other aspects beyond the accuracy, such as response time and the number of interactions, may affect the overall search experience and should be integrated into the framework as well.
Additionally, a comparison between other conversational-based search systems and keyword-based systems with a wider array of metrics could explore the applicability of the preliminary evaluation framework. 
A broader validation by means of a large-scale experiment that represents more demographic groups and industries would foster the generalizability of our framework.

\section{Conclusion}
We propose a semi-automatic approach for a quantitative comparison between traditional keyword-based and semantic search systems, regardless of the respective output formats.
By creating an evaluation dataset of queries (in both natural language and via keywords) with a checklist for content to be found spanning different language styles, lengths, and complexity levels, real-world usage can be simulated. 
We then create ordinal-scaled equivalence classes exhibiting the same retrieval quality based on the checklists. 
%Fiete-final: Was meinst du mit uniquely? Ggf anderes Wort? -> dass es nur zu einer Ä.K. zugewiesen werden kann. -> ah, das kann glaube ich missverstanden werden. Ich würde den ", and uniquely assign outputs to them" Teil vielleicht ganz weglassen. Ich weiß nicht, ob uniquely so gut passt und eigentlich würde ich als leser auch nicht auf die idee kommen mehrere ÄKs zu assignen -> komplett weglassen wird den Schritt nicht explizit erwähnen.  
Investigated (but expandable) criteria are correctness, completeness, and ranking. 
To obtain comparison results, we run statistical tests (e.g., Mann-Whitney U-Test and Vargha-Delaney $A12$).
We validate our preliminary evaluation framework with an industrial case study comparing the Confluence search engine with RAG and its retrieval component separately.
This demonstrates statistically significant improvements in context-aware search over traditional keyword-based methods, as well as small and moderate advantages of RAG and only the retrieval component of RAG over keyword-based methods, respectively.
With its adaptable design, the proposed preliminary framework shows promising potential towards a more objective evaluation between keyword-based and context-aware search methods.
It alleviates the need for costly and time-consuming subjective user studies. At the same time, it ensures a customized evaluation tailored to domain-specific contexts.
%Fiete: Wollen wir darauf enden? Ich glaube ein positiveres Ende wäre besser (diesen Teil vorziehen und darauf enden welche Vorteile das Framework hat und welche Chancen sich ergeben
\section{GenAI Usage Disclosure}
We used ChatGPT to generate code snippets for the statistical tests %(Mann-Whitney U-Test and Vargha-Delaney $A12$ statistic ) 
and the plots. 
The code of the chunking method in RAG, based on the hierarchical structure, is generated by ChatGPT.
We utilized DeepL to translate the resources from German to English. 
The authors carefully reviewed and adjusted all AI-generated content for accuracy and clarity. 
\bibliographystyle{plainnat}  
\bibliography{sample-base}
\end{document}